# Graphene on Silicon Hybrid Field-Effect Transistors

M. Fomin [1,2], F. Pasadas[3], E. G. Marin[3], A. Medina-Rull[3], F. G. Ruiz[3], A. Godoy[3], I. Zadorozhnyi[1], G. Beltramo[4], F. Brings[1,5], S. Vitusevich[1], A. Offenhaeusser[1], and D. Kireev [1,6]

[1] Institute of Bioelectronics (ICS-8/IBI-3), Forschungszentrum Jülich, 52425 Jülich, Germany
[2] Physics Department, University of Osnabrueck, Osnabrueck, Germany
[3] Departamento de Electrónica y Tecnología de Computadores, PEARL Laboratory, Universidad de Granada, Spain
[4] Institute of Biological Information Processing (IBI-2), Forschungszentrum Jülich, 52425 Jülich, Germany
[5] Institute of Materials in Electrical Engineering 1, RWTH Aachen University, Germany
[6] Department of Electrical and Computer Engineering, The University of Texas at Austin, USA

E-mail: s.vitusevich@fz-juelich.de, a.offenhaeusser@fz-juelich.de, kirdmitry@gmail.com

## Abstract

The combination of graphene with silicon in hybrid devices has attracted attention extensively over the last decade. Most of such devices were proposed for photonics and radiofrequency applications. In this work, we present a unique technology of graphene-on-silicon heterostructures and their properties as solution-gated transistors. The graphene-on-Silicon field-effect transistors (GoSFETs) were fabricated exploiting various conformations of drain-source regions doping and channel material dimensions. The fabricated devices were electrically characterized demonstrating hybrid behavior with features specific to both graphene and silicon. Although GoSFET's transconductance and carrier's mobility were found to be lower than in conventional silicon and graphene field-effect transistors (SiFETs and GFETs), it was demonstrated that the combination of both materials within the hybrid channel contribute uniquely to the charge carrier transport. A comprehensive physics-based compact modeling was specifically developed, showing excellent agreement with the experimental data. The model is employed to rationalize the observed hybrid behavior as the theoretical results from the electrostatics and the carrier transport under a drift-diffusion approach show that graphene acts as a shield for the silicon channel, giving rise to a non-uniform potential distribution along it, especially at the subthreshold region. This graphene screening effect is shown to strongly affect the device subthreshold swing when compared against a conventional SiFET due to a non-negligible diffusion current in this operation regime.

# 1. Introduction

Silicon is an iconic material that has been the cornerstone of micro- and nanoelectronics for the last half a century. Silicon-based field-effect transistors (SiFETs) have evolved from being rudimentary and bulky into sub-3nm dimensions [1], [2]. This continuous downsizing has brought numerous advantages but also relevant challenges, coming hand-in-hand with the conception of novel and varied applications [3]–[6]. Among them, biosensing and bioelectronic applications of SiFETs have been recently explored [7], [8] and successfully used as biosensors when specific biomarkers are functionalized on their surface, enabling selective label-free detection [9], [10]. Silicon have also been utilized for numerous *in vitro* recordings of electrogenic cells (cardiac or neural) or even *in vivo* mapping of the whole brain [11], [12]. However, it is known to be a bioresorbable material that degrades over time once immersed in saline, thus, suffering from a limited operation time for *in-vivo* applications [13], [14].

While silicon still dominates the industrial semiconductor scene, a new material has emerged rather recently, transforming materials science: graphene. Graphene, accompanied by other two-dimensional (2D) materials, has already opened new prospects in modern nanoelectronic applications and also holds a great promise for bio- and neuro- applications due to an extraordinary conductivity and good bio-compatibility [15]–[17]. In particular, graphene-based FETs (GFETs) and microelectrode arrays (MEAs), both rigid and flexible, have been reported to successfully interface with electrogenic cells as well as live tissues [18]–[20]. However, the absence of bandgap of graphene results in a large "off" state currents, and the effect of a non-negligible quantum capacitance limits effective out-of-plane electrical coupling to the biomolecules or electrostatic potentials created by the cells [21].

In this work, we propose to merge the two modalities, SiFETs and GFETs, creating a so-called Graphene-on-Silicon FET (GoSFET). Graphene has already been extensively combined with other 2D materials in numerous heterostructures aiming to enhance their electrical properties, as in the case of molybdenum disulfide ($MoS_2$) lateral heterostructures [22], [23]. A much more significant contribution comes from the silicon when graphene field-effect transistors are fabricated on top. In this case, the hybrid channel has already been exploited, for example, to demonstrate ultrasensitive and fast photoresponse [24], [25]; or radiofrequency devices [26], where the bare silicon substrate without an insulating layer was used passively. In our work, on the contrary, both graphene and silicon channels contribute to active charge transfer, as we are seeking to build a more robust device with a high on/off ratio due to silicon, and a high conductivity, transconductance, and environmental stability due to graphene. Experimental findings revealed a hybrid behavior indeed, however the properties of combined channel materials considerably degraded the observed electronic transport properties of the hybrid device, due to a high electrostatic coupling between graphene and silicon . More specifically, the transconductance of GoSFET devices is significantly lower than for the conventional SiFETs and GFETs, and extracted carrier mobility for the graphene channel is reduced to 15.3 $cm^2V^{-1}s^{-1}$ for holes and 23.1 $cm^2V^{-1}s^{-1}$ for electrons [27]–[29].

Finally, in order to rationalize and shed light on the charge transfer phenomena in the hybrid graphene-silicon channel, we developed a comprehensive physics-based model of GoSFETs comprising the electrostatics and the drain current description of the heterostructure. The estimated electrical properties of GoSFETs correlate very well with the experimental results. Due to the placement of graphene directly on top of the silicon, a screening effect is observed, producing a non-uniform charge distribution along the silicon channel in the subthreshold region, which eventually impacts and explains the device performance.

## 2. Results and Discussion

### 2.1 Experimental

The hybrid graphene-on-silicon field-effect transistors are composed of a conventional SiFET where the top-gate metal-insulator interface is substituted by a graphene sheet, electrostatically controlled by a reference electrode immersed into an electrolyte solution on top of the channel. A schematic depiction of a GoSFET can be found in *Fig 1a*. The graphene is laterally contacted with metal stacks of Ti/Au which are on top of the highly doped drain and source edges of the silicon channel forming the drain ($V_D$) and source ($V_S$) electrodes. The electrostatic modulation of the carrier concentration in both graphene and silicon channels is achieved via liquid gating ($V_{LG}$).

We fabricate four wafers with arrays of 32 GoSFET chips, each of them featuring an array 14 transistors. As a base for hybrid device fabrication, we use mildly p-doped SOI wafers ($\rho \approx$ 14-18.9 $\Omega$cm). In order to study the expected interaction between channel material layers, we use different layouts defining the size of the Si and graphene channels. In some cases, graphene has precisely the same size as silicon, while in other samples, it is up to 20 times narrower (see Table S1). First, we use a thermally grown $SiO_2$ hard mask pre-patterned utilizing photolithography and $CHF_3$ plasma etching for silicon nanoribbons formation. Then, following the silicon etching in a solution of tetramethylammonium hydroxide (TMAH), drain-source regions on half of the devices are doped with arsenic and another half with boron. The resulting inversion- ($n^+$−p−$n^+$) and accumulation- ($p^+$−p−$p^+$) mode SiFETs are contacted by metallization using the TiN and Al compound (small squares on the feedlines from the top and bottom part of *Fig. 1b*). Then, Ti/Au metallization of the whole feedlines is performed. Next, the CVD-grown graphene is transferred onto the Si substrate using the "fishing" technique and patterned with oxygen plasma before a second metallization. As previously reported [28], graphene has to be sandwiched between metallization layers to reduce contact resistance. Lastly, the devices are passivated so that only the channel is in direct contact with the liquid electrolyte. An illustrative view of the cross-section of the fabricated devices can be found in *Fig. 1a*, while *Fig. 1b* shows an optical photograph of the original transistor after the passivation step with a visible window opening in the middle.

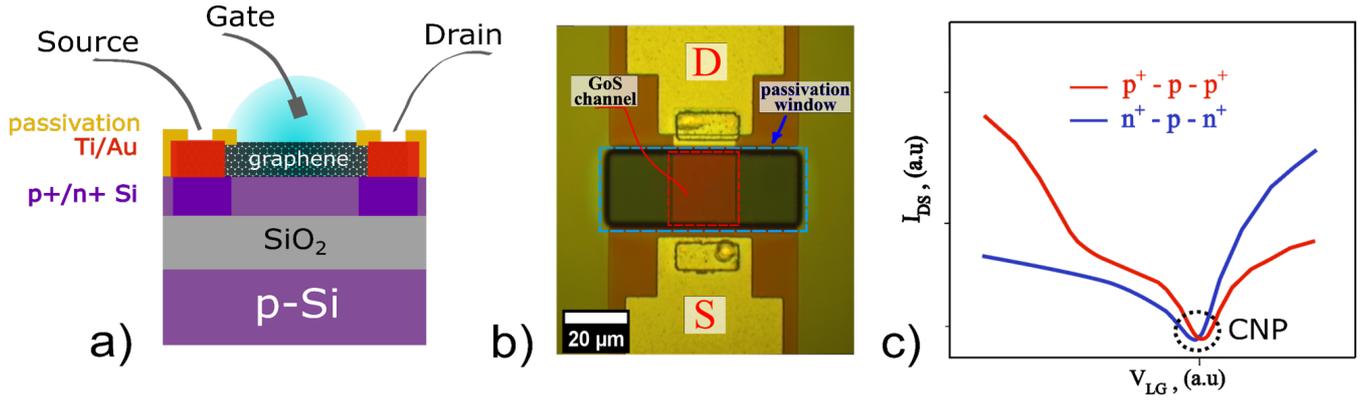

*Fig 1 a) Cross-sectional schematic of the GoSFET device. Silicon is color-coded in violet (or dark violet for high doped regions), metal–red, passivation–yellow. b) Optical photograph of the GoSFET channel (red dashed line), with passivation window (blue dashed line). Passivated feedlines (yellow) on top of the mesa structure (red) contact the combined channel from the bottom and top part of the photograph. c) Illustrative comparison of $I_{DS}$-$V_{LG}$ characteristics for two kinds of GoSFET devices explored in this work, $p^+$–p–$p^+$ (red) and $n^+$–p–$n^+$ (blue), where graphene-like charge neutrality point (CNP) is highlighted. A specific example of the measured I-V curves can be found in Fig. 2a.*

From the illustrative drain current ($I_{DS}$) versus liquid gate voltage ($V_{LG}$) transfer characteristics of GoSFETs shown in *Fig. 1c*, one can notice that hybridized transistors combine typical unipolar silicon behavior with ambipolar graphene. From the experimental data for $p^+$–p–$p^+$ transistors, we hypothesize that a transition between the charge neutrality point (CNP) and silicon-like part is happening where the silicon and graphene channels conductions are balanced. For $n^+$–p–$n^+$ devices its somewhat different and electron conduction of the CNP is covered by dominant silicon conduction without any transition.

The $I_{DS}$-$V_{LG}$ transfer characteristics for one of the $p^+$–p–$p^+$ devices under various drain biases are shown in *Fig. 2a* (see *Fig. 2d* for transfer curves of the $n^+$–p–$n^+$ devices), where the silicon and graphene current dominated regions (as later identified in Section 2.2) are shaded in light blue and yellow, respectively. Here graphene dominated current, as well as current neutrality point are shifted towards positive voltages due to the doping from adsorbates or adsorbates attached to silanol (SiOH) groups at the graphene/Si interface [30].

Also, as can be noticed from *Fig. 2a*, there is a small plateau between $V_{LG}$ of -300 mV and 50 mV. Since the resulting characteristics are the hybrid of SiFET and GFET, we can assume that this part is a transition range where the channel conduction dominated by graphene gets covered by the raising conduction of the silicon part of the channel.

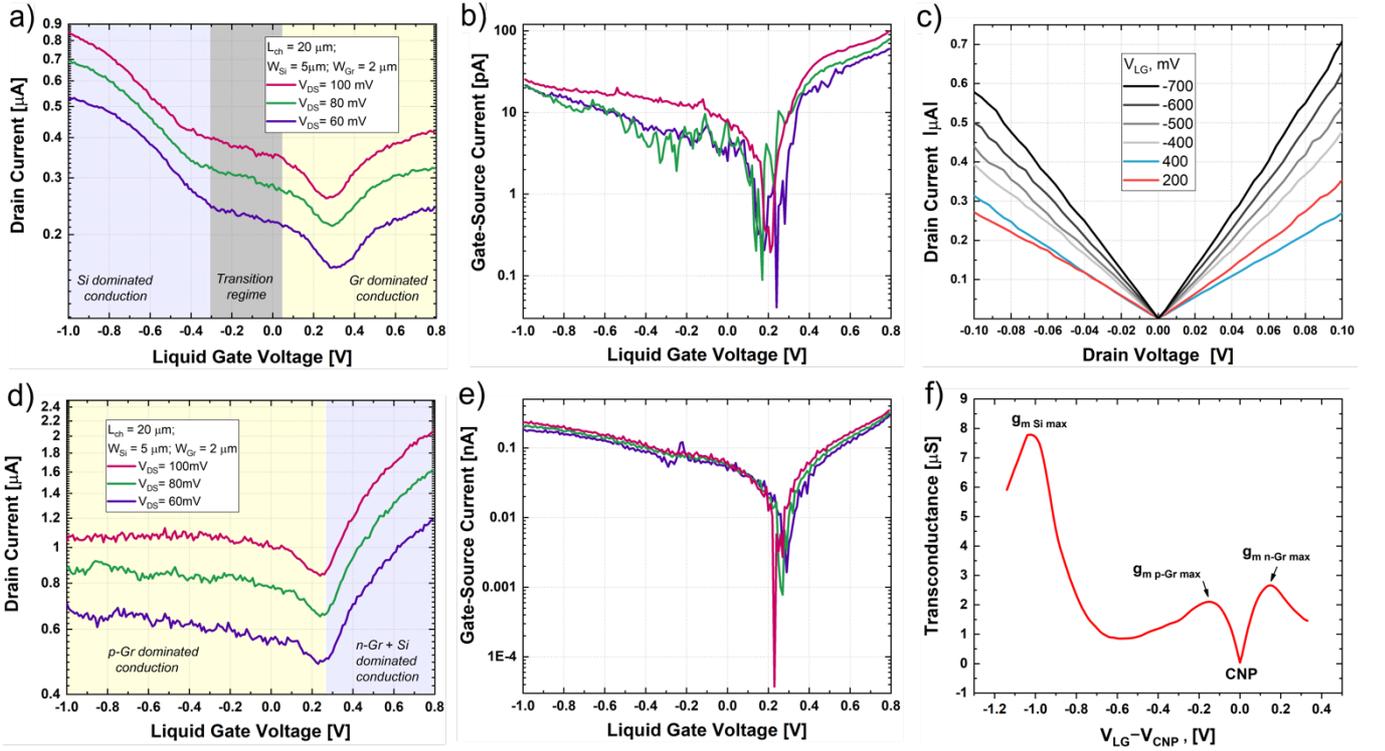

*Fig 2. a)* Transfer characteristics $I_{DS}$-$V_{LG}$ of a $p^+$−$p$−$p^+$ GoSFET under different drain bias voltages and *b)* leakage gate current $I_{GS}$-$V_{LG}$ with corresponding color labeling. *c)* Output characteristics $|I_{DS}|$-$V_{DS}$ of the same device. Here red and blue colors represent current in the linear part of the hole and electron conduction of graphene. The gradual change of the current level in the linear part of the silicon conductive region is shown in shades of black. *d)* Transfer $I_{DS}$-$V_{LG}$ characteristic of one $n^+ - p - n^+$ based GoSFET under various drain voltages, and *e)* corresponding leakage current $I_{GS}$-$V_{LG}$. *f)* Typical transconductance curve for $p+ - p - p+$ devices in absolute value. Here $g_{m\ Si\ max}$ indicates the transconductance maximum of the silicon-like part of the channel; $g_{m\ p\text{-}Gr\ max}$ is the transconductance maximum for hole carriers in graphene-like behavior, and $g_{m\ n\text{-}Gr\ max}$ for electrons.

We also demonstrate that the leakage current for these structures is below ~100 pA for $p^+$−$p$−$p^+$ devices, and below ~300 pA for n−p−n devices (*Figs. 2b,e*), which are negligibly small values compared to the drain-to-source currents. Besides, *Fig. 2c* shows the $|I_{DS}|$-$V_{DS}$ output characteristics of the GoSFET in the transport-dominated regions of both graphene ($V_{LG}$=200mV and 400mV) and silicon ($V_{LG}$ from -700mV to -400mV), evidencing a linear behavior as a function of $V_{DS}$. Next, we extracted the device performance, including the transconductance and mobility, for a total of 20 transistors. The transconductance defines the change in current as a response to change in gate potential, which is the essence of any liquid-gated biosensor; hence it is an important figure-of-merit for GoSFET biosensors. A typical transconductance curve ($g_m = dI_{DS}/dV_{LG}$) for $p^+$−$p$−$p^+$ heterostructures can be found in *Fig. 2f*.

While we can conceptually connect the $g_m$ peaks to a particular transport regime of the hybrid $I_{DS}$-$V_{LG}$ characteristic in Fig 2a, one must keep in mind that the transconductance is not directly explainable in terms of one of the channel materials but as the combined effect of both, as we will demonstrate in Section 2.2. In order to analyze the GoSFETs performance, the two regions with graphene-like and silicon-like behavior are compared. The silicon-like I-V characteristics demonstrates an average maximum transconductance normalized over the bias voltage ($V_{DS}$) in the range of 51.4 µS/V with the standard deviation, $SD$ = 51.8 µS/V. For the graphene-like behavior, we estimated the transconductance separately for both types of carriers. Thus, the mean value of the bias normalized hole transconductance maximum is 20.1 µS/V ($SD$ = 22.7µS/V), while for electrons, it is somewhat higher, 33.6 µS/V ($SD$ = 41.5µS/V). Chartboxes with the statistical data for graphene-like and silicon-like maximum transconductance are provided in *Fig. S1* and in both cases, the values are significantly lower than those achieved in conventional GFETs or SiFETs [27]–[29]. Nevertheless, it should be stressed that these results require a careful interpretation due to the unique nature of the GoSFETs as hybrid devices that cannot be bluntly compared with any of the two.

Furthermore, we show that maximum transconductance of GoSFETs has a certain dependency on the channel geometry (*Fig. S2-4*). A somewhat positive correlation can be tracked between transconductance at the silicon-like behavior part and silicon width in *Fig. S2a*. Also, it should be noted that for graphene-like behavior, both hole and electron transconductances also have a positive trend across silicon width (*Fig. S3-4*). Besides, a positive correlation can be observed between *n*-graphene and the width-to-length silicon ratio (*Fig. S4b*). These dependences are indirect evidence of the hybrid nature of the CNP vertex, what with more details explained later in Section 2.2.

After the transconductance analysis, we estimated the carrier mobility for the graphene-like part of the channel from the DC measurements [28]:

$$\mu = \frac{L}{W} \cdot \frac{g_m}{C_{int} V_{DS}}$$

where *L* and *W* are the channel length and width, respectively, and $\boldsymbol{C_{int}}$ represents the interface capacitance.

The average value of graphene-like hole region maximum mobility is 15.4 cm$^2$V$^{-1}$s$^{-1}$ ($SD$ = 13.8), while the electron region of the characteristics shows a considerably higher value, 23.1 cm$^2$V$^{-1}$s$^{-1}$ ($SD$ = 17.9). In both cases, in order to extract the maximum mobility, the interface capacitance was assumed to be $C_{int} \approx$ *2.1* µF/cm$^2$ according to the theoretical model (see section 2.2). Chartboxes of maximum mobility values for all measured transistors can be found in *Fig. S5*. As well as for the transconductance, GoSFETs graphene-like part carrier mobility is significantly lower compared to conventional devices [27]–[29]. Since W/L ratio is used to calculate carrier mobility, the final values are independent on channel geometry (see *Fig. S6-7*).

During the measurements, it was also revealed that some of the devices were rapidly degrading, which could be easily tracked within several rounds of measurements. We studied this process analyzing the evolution of the transconductance maximum from the first sweep (*Fig. S8*). The upgoing trend in the first measurements can be explained by the elimination of the remanent fabrication residues adhered to the channel materials surface. Then, the transconductance is gradually degraded for both, graphene-like and silicon-like regions, until graphene-like transcondutance becomes undetectable (at this point, the CNP cannot be identified). Our assumption is that graphene is degrading, and once it gets fused, the overall device performance declines. To prove this point, we studied the graphene surface before and after degradation via Raman spectroscopy. The Raman spectra on the newly fabricated GoSFETs demonstrates strongly pronounced G and 2D band signals originated by the monolayer graphene (*Fig. S9*). Simultaneously, Raman spectra acquired from the degraded devices produces only minor signs of the G band peak, while the 2D band peak is almost completely vanished (*Fig. S9a*). From the gained spectra for both cases we managed to develop footprints of the graphene with weighting factors for each pixel (*Fig. S9b-c*). The intensity ratio $I_D/I_G$ of the fitted Gaussians for D and G peaks lies mostly below 1 (blue regions in *Fig. S9b*), indicating good quality low-defect graphene; while the device after many measurement cycles features $I_D/I_G$ ratio greater than 2 indicating the presence of numerous defects in the material [31], [32], and even a transformation of the graphene layer into scattered carbon clusters (yellow regions in *Fig. S9c*), which is indicative of destructed graphene.

## 2.2. Electrostatics of liquid-gated GoSFETs

In order to rationalize the experimental measurements and give insights into the physics at play, we have implemented a comprehensive physics-based electrical model of GoSFETs. The model reproduces to an excellent agreement (maximum average relative error of 7.5%) the experimental electrical readouts and provides an in-depth description of the device electrical behavior supporting the interpretation of the measurements.

As already mentioned in the experimental section, the electrostatic modulation of the carrier concentration in the hybrid GoSFET channel is achieved *via* the reference electrode ($V_{LG}$) immersed into the aqueous solution. In order to understand the dependence of the carrier density in graphene and silicon with the reference electrode potential, we first deal with the charge distribution at the graphene/electrolyte interface [33]. In particular, it is crucial to properly model the capacitance at such interface so to accurately relate the charge carrier density induced in both graphene and silicon with the electrode potential. The interfacial capacitance can be split into three contributions, as represented in the equivalent capacitive circuit of GoSFETs shown in *Fig. 3a*: a double layer capacitance ($C_{DL}$), a Stern capacitance ($C_{Stern}$), and a gap capacitance ($C_{gap}$). $C_{DL}$ accounts for the electrical double layer that appears at the electrolyte/graphene interface [34] and ranges from a few µF/cm$^2$ to a few hundreds of µF/cm$^2$ depending on the metal electrodes or the ionic concentration of the electrolyte [34], [35]. $C_{Stern}$ models the region depleted of ionic charges close to the surface [36]–[38], with values also varying among tens of µF/cm$^2$ [39]. Finally, $C_{gap}$ considers the hydrophobic nature of the graphene surface and the consequent changes in the electrolyte close it [40]: as demonstrated by molecular dynamics simulations [38], the density of water decreases strongly at the surface, resulting in a so-called hydrophobic gap between the solid and the electrolyte. In this gap, the effective dielectric constant is smaller than in bulk water, resulting in a large potential drop at the interface. A hydrophobic gap of 0.31 nm and a dielectric constant of 1 are considered [27], [28], resulting in $C_{gap} \approx 2.1$ µF/cm$^2$.

The series combination of $C_{DL}$, $C_{Stern}$, and $C_{gap}$ results in an equivalent interfacial capacitance ($C_{int}$) which is dominated by $C_{gap}$, with a capacitance value around one order of magnitude lower than $C_{DL}$ and $C_{Stern}$ given that the ionic concentration, $i_0$, is at least 10 mM. Thus, the electrode-electrolyte-hydrophobic gap heterostructure acts as an effective capacitance $C_{int} = (C_{DL}^{-1} + C_{Stern}^{-1} + C_{gap}^{-1})^{-1} \approx C_{gap}$, where an effective potential, $-\Psi_0$, drops across the electrolyte/graphene interface [28], resulting in an effective gating $V_G = V_{LG} + \Psi_0$. The potential $\Psi_0$, according to the site-binding theory [41], is dependent on pH, ionic concentration, the density of surface ionizable sites and dissociation constants [42], [43]. Here $\psi_0$ is considered as a gate offset bias as the electrolyte characteristics are not modified during the experimental measurements.

The next region in the device heterostructure is the graphene-silicon interface. In order to model it, we consider the formation of a interfacial silicon-graphene dipole layer within an equilibrium separation distance, $t_{eq}$ = 0.6 nm [44], [45] and a dielectric constant of 1 (resulting in $C_{dip} \approx 1.5$ µF/cm$^2$). In this regard, the 1D electrostatics is analyzed by solving the Gauss' law (see Methods) across the electrode-electrolyte-hydrophobic gap-graphene-dipole layer-silicon heterostructure shown in *Fig. 3b*. The electrostatics of the GoSFET is then described using the equivalent capacitive circuit depicted in *Fig. 3a*, where $C_q = \partial Q_{net}/\partial V_c$ represents the quantum capacitance of graphene [46], while $C_{si} = -\partial Q_{si}/\partial \psi_s$ is the intrinsic silicon capacitance [47], accounting for the 2D and 3D finite density of states of graphene and silicon, respectively.

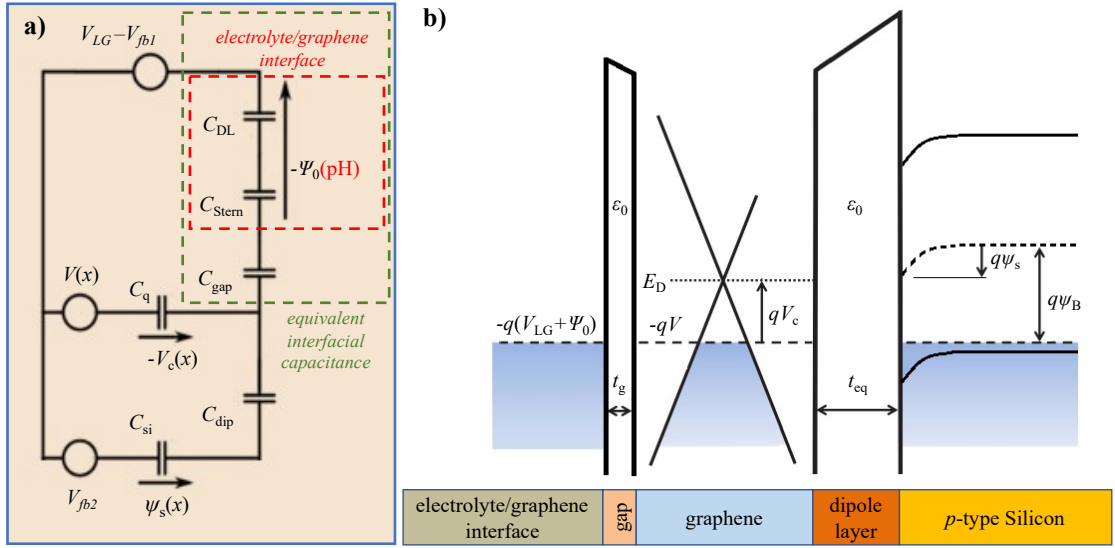

*Fig. 3 a)* Equivalent capacitive circuit of an electrolyte-gated GoSFET. *b)* Sketch of the band diagram of the electrolyte-hydrophobic gap-graphene-dipole layer-silicon heterostructure in equilibrium ($V_{LG} = 0V$). At the graphene channel, $E_D$ is the Dirac energy, $E_D = -q(V-V_c)$, $-qV$ is the graphene Fermi level, and $V_c$ is the graphene chemical potential. At the silicon channel, $q\psi_B$ is the difference between the Fermi level and the intrinsic Fermi level; and $\psi_s$ is the surface potential.

A study of the gate voltage dependence of the relevant capacitances of the accumulation and inversion GoSFETs is addressed in the Supporting Information (*Fig. S10b and Fig. S12b*). It is worth noting that Eq. (1a) (see Methods) resembles the electrostatics of a dual-gated GFET [48] where the role of the top (back) insulator is played by the hydrophobic gap (dipole layer) and the effective top (back) gate voltage is $V_{LG} + \Psi_0 - V_{fb1}$ ($\psi_s + V_{fb2}$). In addition, Eq. (1b) solves the electrostatics of a common SiFET [47] but gated by the graphene potential $V-V_c$ and substituting the insulator by a dipole layer. This way, the electrostatics of a hybrid GoSFET can be understood as the self-consistent solution of both GFET and SiFET devices. The analytical equations used to solve Eq. (1) can be found in *Supplementary Note S1*.

## 2.3 Drift-diffusion transport along the graphene-silicon channels

Given the length of the fabricated devices ranges from 5 to 20 μm, it is reasonable to assume that the mean free path of carriers is much shorter than the channel length, and the drift-diffusion theory is the appropriate framework to describe the electrical transport. In this regard, the drain current can be written in the form $I_{DS} = WQ_t(x)\mu(x)dV(x)/dx$, where $W$ is the gate width; $Q_t$ is the mobile charge density; and $\mu$ is the carrier mobility. We assume that graphene and silicon behave as two independent transport channels, *i.e.*, the flow of current or tunneling between the inversion (accumulation) channel formed at the silicon surface and at the graphene sheet is negligible in comparison with the longitudinal current along each of them.

In graphene, ambipolar $Q_t$ can be expressed as a quadratic polynomial dependent on the chemical potential $V_c$ [49], [50], while symmetrical electron and hole mobility are considered independent of the applied electric field, carrier density or temperature. On the other hand, $Q_t$ in silicon arises from the inversion in the $n^+$–p–$n^+$ GoSFET (or accumulation in the $p^+$–p–$p^+$ GoSFET) channel charge density; with mobility $\mu_{e,Si}$ ($\mu_{h,Si}$) for electrons (holes). Analytical expressions for the calculation of the graphene and silicon currents for both inversion/accumulation regimes of GoSFETs are provided in *Supplementary Note S2*.

## 2.4. Theoretical interpretation of experimental data

TABLE I. $p^+$-p-$p^+$ GoSFET model parameters

| Graphene | | Silicon | | | |
|---|---|---|---|---|---|
| $L$ (μm) | 20 | $L$ (μm) | 20 | $C_{DL}$ (μF/cm$^2$) | 23 |
| $W_g$ (μm) | 2 | $W_s$ (μm) | 2 | $C_{Stern}$ (μF/cm$^2$) | 20 |
| $V_{fb1}$ (V) | 0.3 | $V_{fb2}$-$V_D$ (V) | -0.23 | $C_{gap}$ (μF/cm$^2$) | 2.86 |
| $\mu_g$ (cm$^2$/Vs) | 10 | $\mu_{h,Si}$ (cm$^2$/Vs) | 400 | $C_{dip}$ (μF/cm$^2$) | 1.48 |
| $\sigma_{pud}/q$ (cm$^{-2}$)* | 1.7×10$^{13}$ | $N_A$ (cm$^{-3}$) | 4×10$^{15}$ | $T$ (K) | 300 |

* $\sigma_{pud}$ is the residual charge density due to electron-hole puddles (see *Supplementary Note S2*).

We exploited the implemented model in order to explain the physics at play in the experimental device realizations. To this purpose, we employed the electrical and physical parameters collected in Table I as extracted for the $p^+$-p-$p^+$ GoSFET. In particular, the experimental device of choice comprised a 20 μm-long and 2 μm-wide graphene sheet covered on top with the electrolyte, with an offset voltage $V_{fb1}$ = 0.3V, meaning that it is *p*-type in equilibrium. Notably, the residual charge density is relevant (1.7×10$^{13}$ cm$^{-2}$), and the mobility is low ($\mu_g$ = 10 cm$^2$/Vs), meaning that the graphene sheet is highly contaminated and the carriers suffer from frequent scattering reducing their mobility. The former can be produced by charged impurities and a highly corrugated silicon surface [51], [52], while the latter might be due to a high residual carrier concentration in the channel [53], remote phonons, and the high electrostatic coupling given that the graphene sheet is sandwiched between the silicon surface charge and the electrical double layer originated at the electrolyte interface [54]. The silicon channel has equal dimensions (20 μm-long and 2 μm-wide), and it consists of a *p*-type substrate with acceptor concentration $N_A = 4 \times 10^{15}$ cm$^{-3}$ and hole mobility of $\mu_{h,Si} = 400$ cm$^2$/Vs. The electrolyte is a 1xPBS, so a $C_{DL}$ of 23 μF/cm$^2$ is assumed (calculated from site-binding theory [42]), while $C_{Stern}$ of 20 μF/cm$^2$ was taken [39].

*Fig. 4a* shows the simulated transfer characteristics of the accumulation GoSFET together with the experimental measurements. The model is able to reproduce to an excellent agreement the experimental results (maximum average relative error of 7.5%) with the shape of the experimental data showing a hybrid behavior where the point of minimum conductivity ($V_{MC}$) is highlighted. Some deviation between the simulated and measured data is observed around $V_{LG} \sim -0.5$V. Specifically, we observe from measurements a more pronounced change in the current at such gate voltage. This deviation is originated by a theoretical overestimation of the transport charge density. This issue is thoroughly commented in *Supplementary Note S3* but it does not significantly affect the explanation of the physical operation of the device.

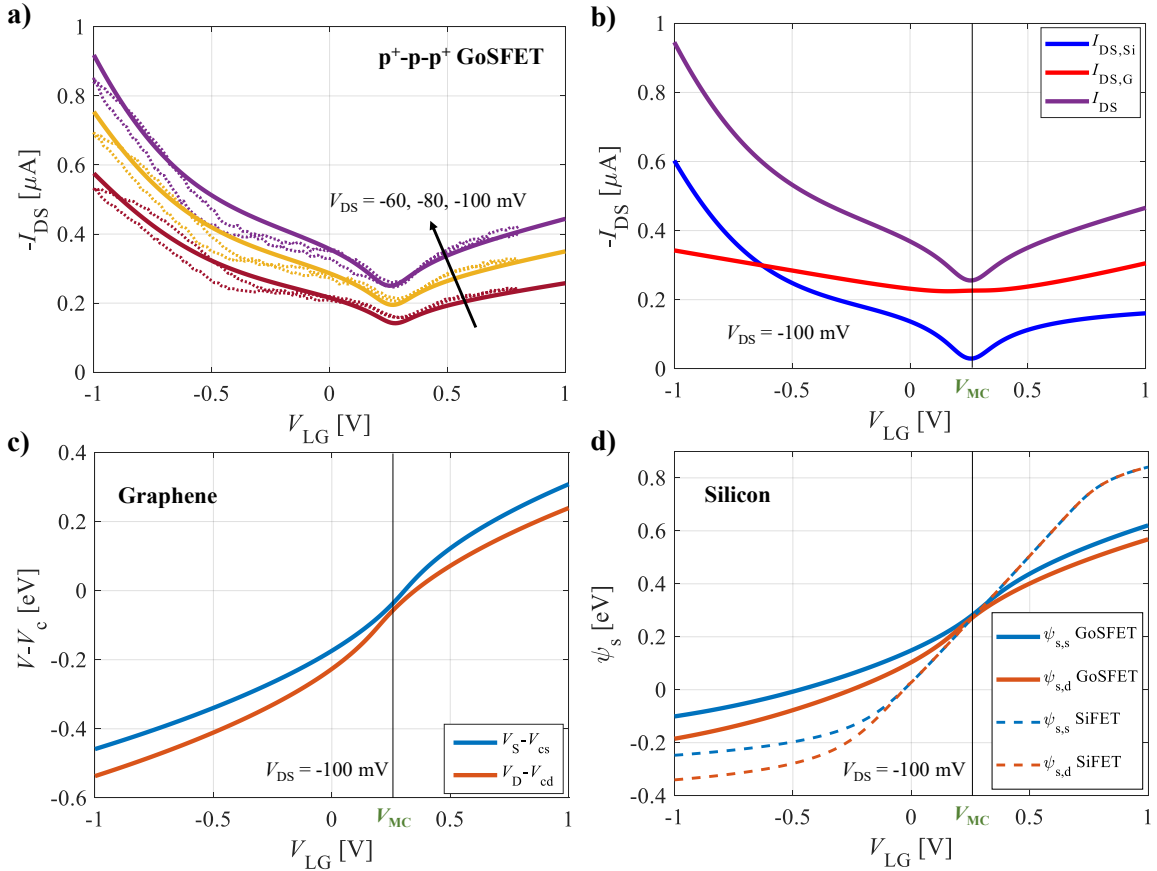

***Fig. 4** a) Transfer characteristics of the accumulation $p^+$-$p$-$p^+$ GoSFET described in Table I for three different drain biases. Simulations are plotted with solid lines and experimentally measured data (forward and backward sweeps) with dots. **b)** Theoretical drain current of the accumulation GoSFET for $V_{DS} = -0.1V$ (purple) split into the graphene (red)* *and silicon (blue) current contributions. A vertical black solid line marks the gate bias that achieves the minimum conductivity, labeled as $V_{MC}$.* ***c)*** *Modulation of the graphene electrostatic potential $V-V_c = E_D/q$ at the source (blue) and drain (orange) ends with a drain voltage of $V_{DS} = -0.1V$.* ***d)*** *Modulation of the silicon surface potential at the source (blue) and drain (orange) edges in the accumulation GoSFET (solid lines) and a SiFET (dashed lines) with $V_{DS} = -0.1V$.*

The contributions of graphene and silicon channels to the theoretical current of the p⁺-p-p⁺ GoSFET for $V_{DS}$ = -0.1V are shown in *Fig. 4b*. The minimum conductivity of the accumulation GoSFET occurs at a gate bias $V_{MC}$ = 0.26V. As can be expected, the current at the graphene channel ($I_{DS,G}$) is hardly modulated by the gate voltage due to the high residual concentration (the mobile charge in graphene is hardly tunable in the vicinity of the Dirac voltage in this scenario). In *Fig.4c* the modulation of the graphene electrostatic potentials at the drain and source edges are shown. At the Dirac voltage, half of the graphene channel is filled with electrons and half is filled with holes, meaning that the Fermi level crosses the Dirac energy at the middle of the graphene channel. For liquid gate voltage higher and lower than CNP voltage, the flow of electrons and holes, respectively, becomes predominant [55]–[58].

The silicon-based charge carriers, on the contrary, are considerably modulated by the gate voltage and the current flowing through the silicon channel (*Fig. 4b*), and they are the main responsible of the overall behavior observed in the current of the GoSFET. Indeed, the vertex of the V-shaped transfer curve of the GoSFET is mainly produced by the current flowing through the silicon channel (*Fig. 4b*). The surface potentials at the drain ($\psi_{s,d}$) and source ($\psi_{s,s}$) edges of the accumulation GoSFET (solid lines) are shown together with these quantities for a SiFET counterpart. According to *Fig. 4d*, $\psi_{s,d}$ and $\psi_{s,s}$ get close at $V_{MC}$, meaning that the transport charge can be considered uniform along the channel, therefore, minimizing the diffusion component of the current. For $-0.5V < V_{LG} < V_{MC}$ and $V_{LG} > V_{MC}$, the surface potentials start to split and thus, both diffusion and drift currents increase. Finally, for $V_{LG} < -0.5V$, when $\psi_s < 0$ holes accumulate in the channel, showing a large increment in the drift current (see details in *Supplementary Note S3*). The surface potentials of a SiFET without considering the graphene layer (see *Fig. 4d*), exhibit usual behavior of an accumulation SiFET. In this case, when $\psi_{s,d}$ and $\psi_{s,s} > 0$, they are almost identical and the diffusion current is negligible. For $\psi_s < 0$, the channel enters in accumulation and the surface potentials start to separate, therefore, increasing drift and diffusion currents.

A complete analysis of the inversion n⁺-p-n⁺ GoSFET is also provided in *Supplementary Note S4*. It must be highlighted that the same parameters collected in Table I have been used for the inversion n⁺-p-n⁺ GoSFET, except the values of the mobilities ($\mu_g$ = 40 cm²/Vs; $\mu_{e,Si}$ = 225 cm²/Vs), the residual charge density due to electron-hole puddles ($\sigma_{pud}/q$ = 5.4×10¹² cm⁻²) and the offset bias ($V_{fb2}-V_D$ = -0.41V) and the agreement between simulation and measurements is excellent (maximum average relative error of 6%) even though each type of GoSFET (accumulation and inversion) shows a quite different transfer characteristic shape (*Fig. 1c*).

## Conclusions

In summary, graphene-on-silicon field effect transistors fabricated by a novel hybrid technology have demonstrated unique behavior where both channel materials contribute to the electronic transport. Resulting structures can be utilized as a biosensor where eventually graphene and silicon can be functionalized for complex analytics separately. In spite of the fact that GoSFETs transconductance and carrier mobility are significantly lower compared to the conventional GFETs and SiFETs, several unique findings were revealed. In particular, contrary to the behavior registered in a conventional SiFET, which is controlled by a constant gate potential along the channel, the GoSFET is characterized by a non-uniform graphene potential that controls the silicon conductivity. Concomitantly, a strong electrostatic coupling is produced between carriers in both channels, an effect addressed in our theoretical analysis and explained by the presence of a non-negligible diffusion current in the subthreshold region originated by the splitting of the drain and source surface potential. Due to this non-uniform gating effect, caused by the graphene layer, the current in the silicon channel in a GoSFET can hardly be switched off. However, this hybrid behavior is registered and correctly described in the frame of our comprehensive theoretical model. Beyond the insights and explanations of the experimental findings here obtained, the model can be used to analyze other kinds of 2D material based heterostructures.

## 3. Methods

**Device fabrication**

Graphene-on-Silicon heterostructures were fabricated on <100> 4-inch silicon-on-insulator (SOI) wafers provided by SOITEC, France. The active silicon layer was 50 nm thick with 145 nm of buried oxide. In the first step, the thermal oxidation of the top silicon layer in the dry oxygen atmosphere was performed (940ºC, 45 min.). Then utilizing e-beam photolithography, the $SiO_2$ hard mask was patterned by anisotropic reactive ion etching (RIE) in $CHF_3$ plasma. The finite silicon nanoribbon shape was transferred to the active silicon layer by dipping wafers into a 5% TMAH water solution at 80ºC for 15 seconds. After stripping off the hard mask with 1% HF (3 min 30 sec), in order to create good ohmic contacts, ion implantation of drain-source terminals is carried out. Depending on the structure, boron (6 keV, $10^{15}$ cm$^{-2}$) and arsenic (8 keV, $10^{14}$ cm$^{-2}$) were used to get highly doped regions p- or n-type, correspondingly. Then, wafers are annealed for dopant activation: 5 seconds at 1000 ºC for boron and 30 seconds at 950 ºC for arsenic implantation. The fabrication layouts of GoSFETs are designed with the possible back gate control. For this reason, we performed etching through the buried oxide by a buffered oxide etch (BOE) for 70 seconds. As a protecting layer pre-patterned, AZ 5214 E photoresist was used. Metallization of the silicon part of the transistors was done in two steps. First, the stack of 5 nm TiN and 200 nm of Al were deposited onto drain-source contacts with the subsequent annealing in forming gas atmosphere ($N_2$:$H_2$ = 10:1 at 450 ºC for 1 min.). This step is essential for the creation of good ohmic contact with the structure. After, metallization of feedlines by 10 nm Ti and 60 nm Au compound was carried out. Finally, after creating inversion (n+ − p − n+) or accumulation (p+ − p − p+) mode SiFETs, we go to the graphene part of the technology. A monolayer of CVD-grown graphene used in this study was provided by Graphenea, Spain. For "fishing" technique transfer, we used a PMMA photoresist as support to transfer graphene on top of the wafers. The PMMA residues are removed by subsequent immersion of the wafers into hot acetone and propanol (60 ºC, 1h. each). Next, patterning of graphene was done using oxygen plasma (1min., 100W). Then, the second metallization with another Ti/Au (10nm/60nm) stack is performed to sandwich graphene. At the last step, each wafer is covered by polyimide for passivation. After photolithography, the passivation layer is annealed and wafers are diced onto chips.

## Characterization

After fabrication stage, devices were fasten to the chip carriers and encapsulated. Characterization was carried out utilizing a Keithley 4200 SCS semiconductor parameter analyzer. The gate potential ($V_{LG}$) was swept against Ag/AgCl pellet electrode from -1 V to 0.8 V. For the liquid gating, a physiologically close 150 mM phosphide buffer solution (PBS) with pH 7.4 was used. The drain-source potential ($V_{DS}$) was altered from 20 mV to 100 mV or from 50 mV to 500 mV with steps of 20 mV or 50 mV respectively.

## Raman spectroscopy

Confocal Raman spectroscopy for GoSFETs was performed using a Witec 300 Alpha R equipped with a Mitutoyo M Plan Apo SL 100×/0.55. objective. The spectra were taken with an excitation laser wavelength of 532 nm and power of 0.3 mW before the objective to avoid damage to the sample. As a laser source applied through a 100 μm single-mode glass fiber a single-mode frequency-doubled Nd:YAG laser was used. The excitation line was isolated from the Raman signal via an edge filter. As a pinhole for Raman confocality served 50 μm multimode fiberglass. Additionally, Raman setup was equipped with Newton Andor EMCCD camera with 1600×200 pixels and a holographic grating of 600 lines/mm. The data processing was performed by cluster analysis and non-negative matrix factorization.

## Electrostatics of the GoSFET

1D electrostatics across the electrode-electrolyte-hydrophobic gap-graphene-dipole layer-silicon heterostructure:

$$\begin{cases} Q_{net}(x) = -C_{int}\left(V_{LG} + \Psi_0 - V_{fb1} + V_c(x) - V(x)\right) - C_{dip}\left(V_{fb2} + \psi_s(x) + V_c(x) - V(x)\right) & \text{(1a)} \\ Q_{si}(x) = -C_{dip}\left(V(x) - V_c(x) - V_{fb2} - \psi_s(x)\right) & \text{(1b)} \end{cases}$$

where $Q_{net}$ is the graphene overall net sheet charge density; $Q_{si}$ is the charge induced in the silicon channel; $V$ is the graphene quasi-Fermi level and must fulfill the boundary conditions: (i) $V(x=0)=V_S$ (source voltage) at the source end; (ii) $V(x=L)=V_D$ (drain voltage) at the drain edge, where $x$ is the transport direction and $L$ is the gate length; $V_c$ is the graphene chemical potential (related to the shift of the Fermi level with respect to the Dirac energy [48]); and $\psi_s$ is the silicon surface potential [47]. Furthermore, $V_{fb1} = \phi_m - \phi_g - Q_{d1}/C_{int}$ and $V_{fb2} = \phi_g - \phi_{si} - Q_{d2}/C_{dip} + V_D$ are the flat-band voltages that comprise metal ($\phi_m$) and graphene ($\phi_g$) work-functions; possible additional charges ($Q_{d1}$ and $Q_{d2}$) due to impurities, doping, etc., at the electrolyte-graphene interface and the graphene-silicon interface; the silicon work-function defined $\phi_{si} = \chi_{si}+E_G/(2q)+\psi_B$, where $\chi_{si}$ and $E_G$ are the silicon electron affinity and band gap; and $\psi_B = (k_BT/q)\ln[N_A/n_i]$ is the difference between the Fermi level and the intrinsic Fermi potentials at the silicon channel [47] (with $k_B$ the Boltzmann constant, $T$ the temperature, $q$ the electron elementary charge, $N_A$ the acceptor concentration at the $p$-type silicon substrate and $n_i$ the silicon intrinsic carrier concentration). Finally, the last term of $V_{fb2}$ embraces the $V_D$ dependence of the shift of the Dirac voltage of graphene due to traps at the graphene-silicon interface [59]. As aforementioned, since the electrolyte properties are not modified, $\Psi_0$ is constant in Eq. (1a) and could be incorporated into $V_{fb1}$ as a correction of the flat-band condition.

## Acknowledgments

This work was supported by MCIN/AEI/10.13039/501100011033 through the project PID2020-116518GB-I00, by FEDER/Junta de Andalucía-Consejería de Transformación Económica, Industria, Conocimiento y Universidades through the Projects B-RNM-375-UGR18 and PY20_00633. UGR members also acknowledge the support by the European Union's Horizon 2020 Framework Programme for Research and Innovation through the Project Wearable Applications Enabled by Electronic Systems on Paper (WASP) under Contract 825213. Francisco Pasadas acknowledges funding from PAIDI 2020 and European Social Fund Operational Programme 2014–2020 no. 20804. A. Medina-Rull acknowledges the support of the MCIN/AEI/PTA grant, with reference PTA2020-018250-I.

# Supplementary Information for:

# Graphene on Silicon Hybrid Field-Effect Transistors

**M. Fomin [1,2], F. Pasadas[3], E. G. Marin[3], A. Medina-Rull[3], F. G. Ruiz[3], A. Godoy[3], I. Zadorozhnyi[1], G. Beltramo[4], F. Brings[1,5], S. Vitusevich[1], A. Offenhaeusser[1], and D. Kireev [1,6]**

[1] Institute of Bioelectronics (ICS-8/IBI-3), Forschungszentrum Jülich, 52425 Jülich, Germany
[2] Physics Department, University of Osnabrueck, Osnabrueck, Germany
[3] Departamento de Electrónica y Tecnología de Computadores, PEARL Laboratory, Universidad de Granada, Spain
[4] Institute of Biological Information Processing (IBI-2), Forschungszentrum Jülich, 52425 Jülich, Germany
[5] Institute of Materials in Electrical Engineering 1, RWTH Aachen University, Germany
[6] Department of Electrical and Computer Engineering, The University of Texas at Austin, USA

E-mail: s.vitusevich@fz-juelich.de, a.offenhaeusser@fz-juelich.de, kirdmitry@gmail.com

***Supplementary Table S1.*** Materials dimensions on transistors for the first half of the chips (left) and second half of the chips (right). Here L stands for length and W for width.

| Transistor No | Graphene | | Si nanoribbon | | Transistor No | Graphene | | Si nanoribbon | |
|---|---|---|---|---|---|---|---|---|---|
| | L, µm | W, µm | L, µm | W, µm | | L, µm | W, µm | L, µm | W, µm |
| 1 | 20 | 1 | 20 | 5 | 1 | 5 | 5 | 5 | 5 |
| 2 | 20 | 2 | 20 | 5 | 2 | 5 | 10 | 5 | 10 |
| 3 | 20 | 1 | 20 | 10 | 3 | 5 | 15 | 5 | 15 |
| 4 | 20 | 2 | 20 | 10 | 4 | 5 | 20 | 5 | 20 |
| 5 | 20 | 4 | 20 | 10 | 5 | 10 | 5 | 10 | 5 |
| 6 | 20 | 1 | 20 | 20 | 6 | 10 | 2 | 10 | 2 |
| 7 | 20 | 2 | 20 | 20 | 7 | 10 | 10 | 10 | 10 |
| 8 | 20 | 4 | 20 | 20 | 8 | 10 | 15 | 10 | 15 |
| 9 | 10 | 1 | 10 | 20 | 9 | 10 | 20 | 10 | 20 |
| 10 | 10 | 2 | 10 | 20 | 10 | 20 | 2 | 20 | 2 |
| 11 | 10 | 4 | 10 | 20 | 11 | 20 | 5 | 20 | 5 |
| 12 | 5 | 1 | 5 | 20 | 12 | 20 | 10 | 20 | 10 |
| 13 | 5 | 2 | 5 | 20 | 13 | 20 | 15 | 20 | 15 |
| 14 | 5 | 4 | 5 | 20 | 14 | 20 | 20 | 20 | 20 |

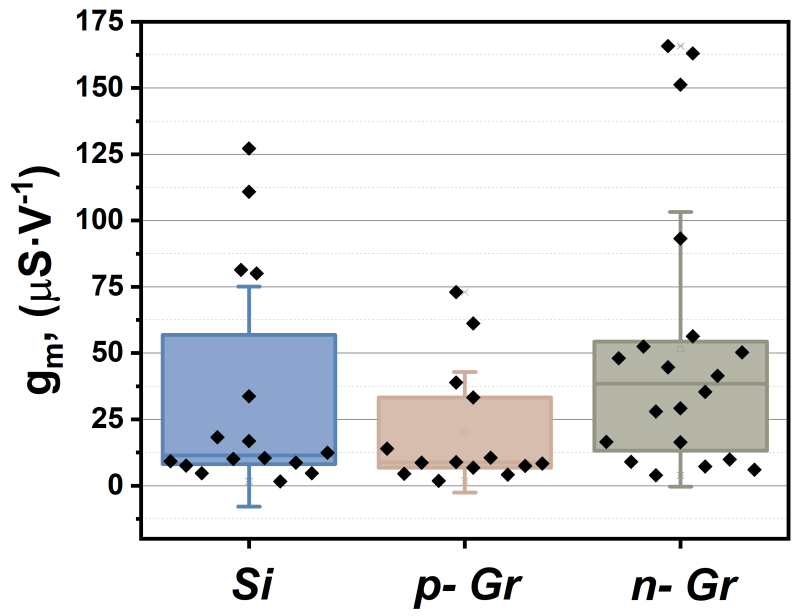

***Supplementary Figure S1.*** *Transconductance maximum statistics from N=20 $p^+ - p - p^+$ transistors. Some data points are missing since for some of the devices it was not possible to extract the transconductance maximum due to the noise.*

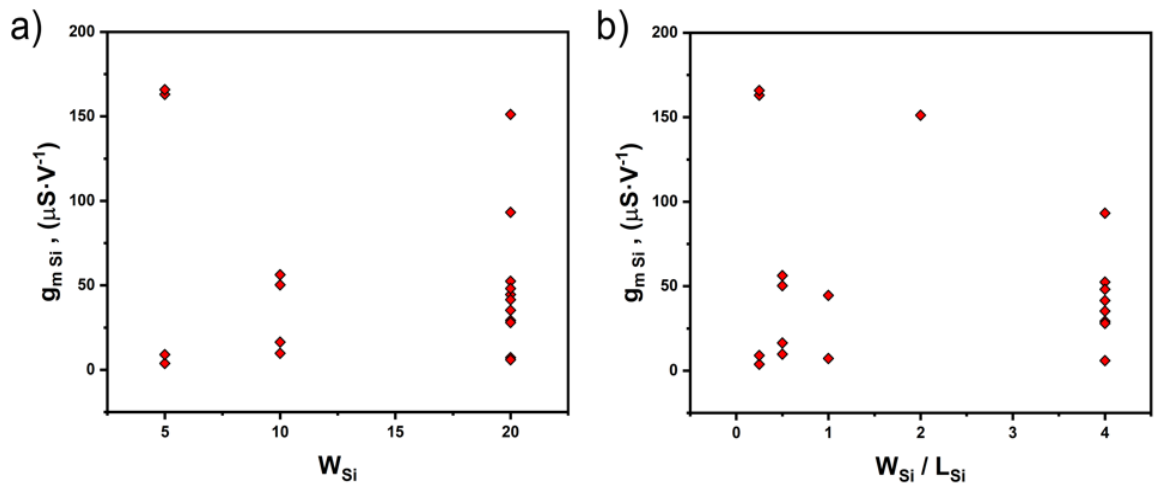

***Supplementary Figure S2.*** *Transconductance maximum dependencies for the silicon section of the characteristics versus (a) width of the silicon channel and (b) width-to-length ratio of the silicon channel.*

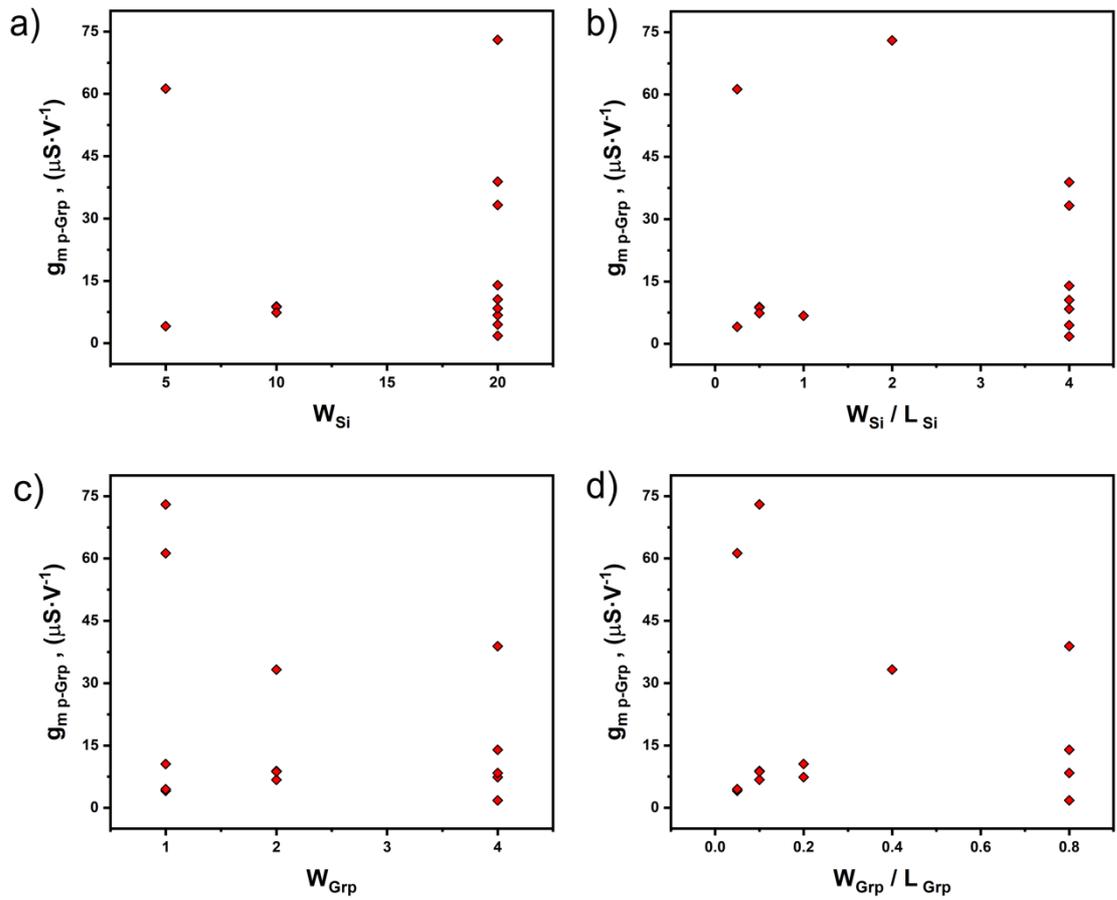

***Supplementary Figure S3.*** *Transconductance maximum dependencies for hole carriers in the graphene part of the characteristics versus (a) width of the silicon channel; (b) width-to-length ratio of the silicon; (c) width of graphene and (d) width-to-length ratio of graphene.*

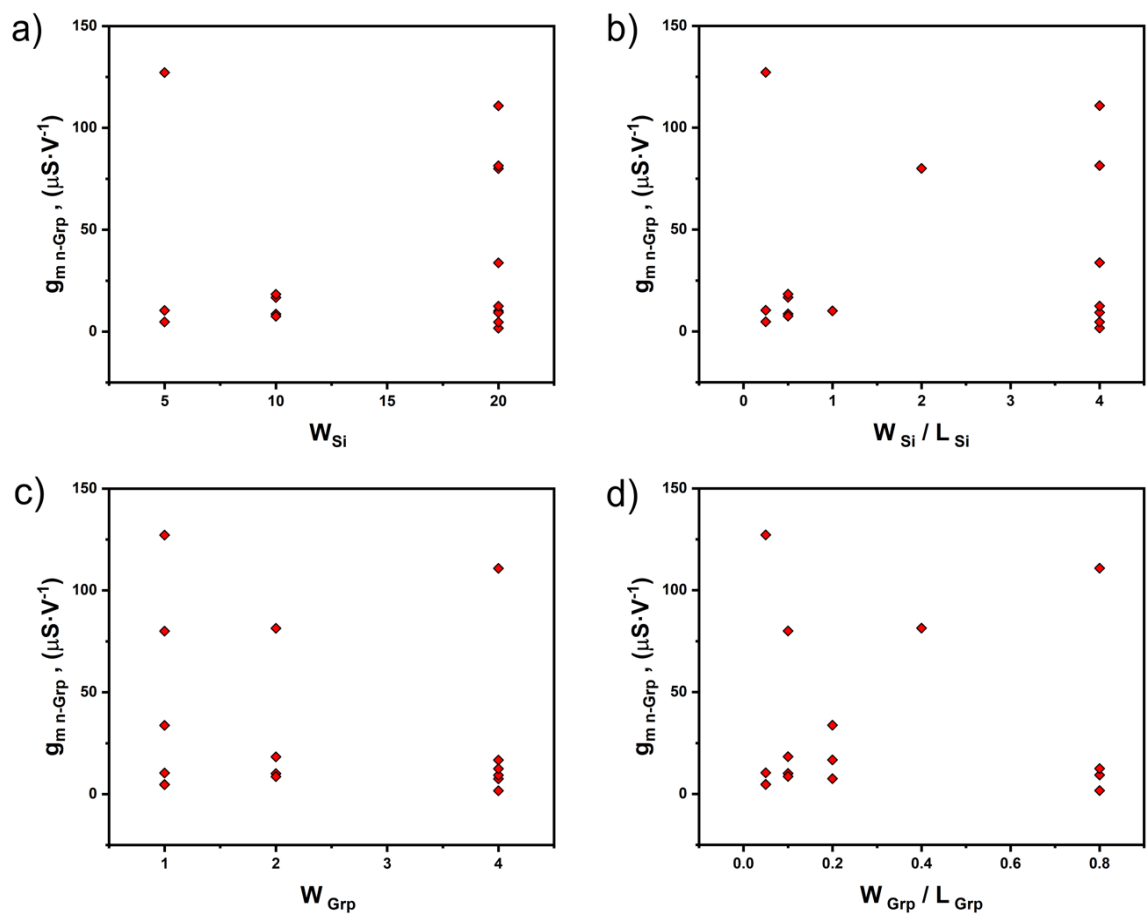

***Supplementary Figure S4.*** Transconductance maximum dependencies for electron carriers in the graphene part of the characteristics versus (a) width of the silicon channel; (b) width-to-length ratio of the silicon; (c) width of graphene and (d) width-to-length ratio of graphene.

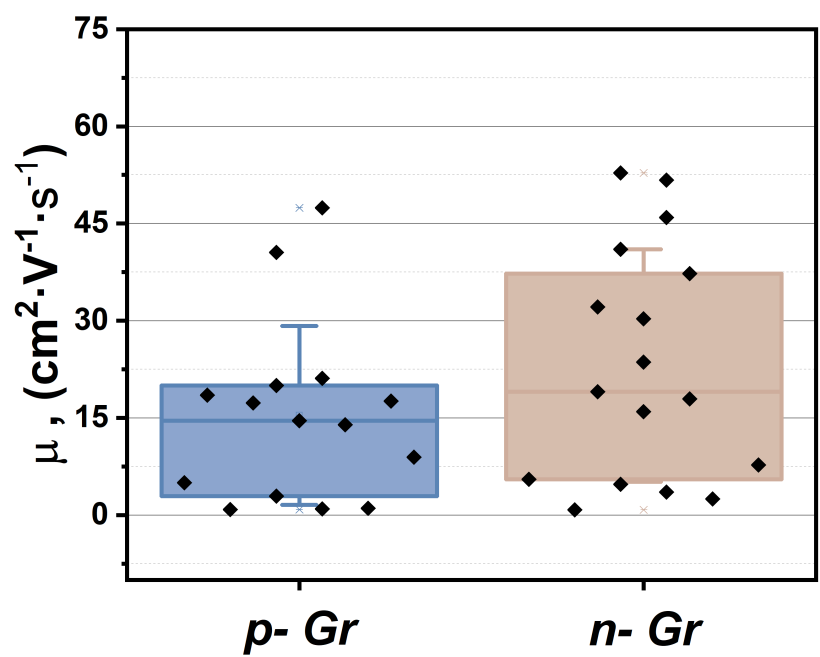

***Supplementary Figure S5.*** *Hole and electron maximum mobility comparison, estimated for the graphene part of the characteristics of GoSFETs.*

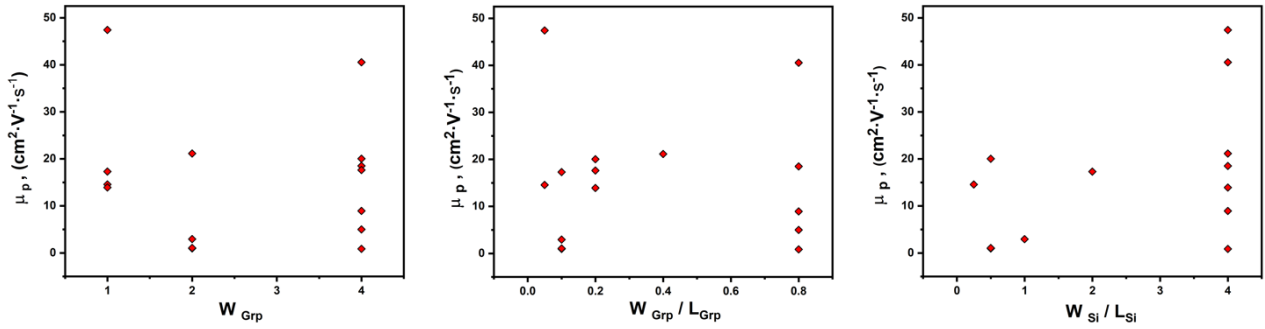

***Supplementary Figure S6.*** *Hole mobility maximum for the graphene section of the characteristics versus (a) width of graphene channel; (b) width-to-length ratio of graphene; (c) width-to-length ratio of the silicon.*

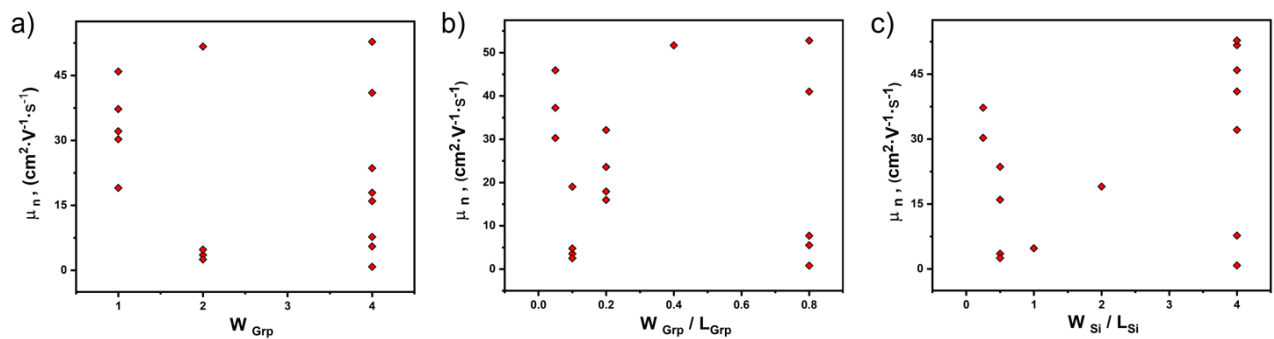

***Supplementary Figure S7.*** *Electron mobility maximum for the graphene section of the characteristics versus (a) width of graphene channel; (b) width-to-length ratio of graphene; (c) width-to-length ratio of the silicon.*

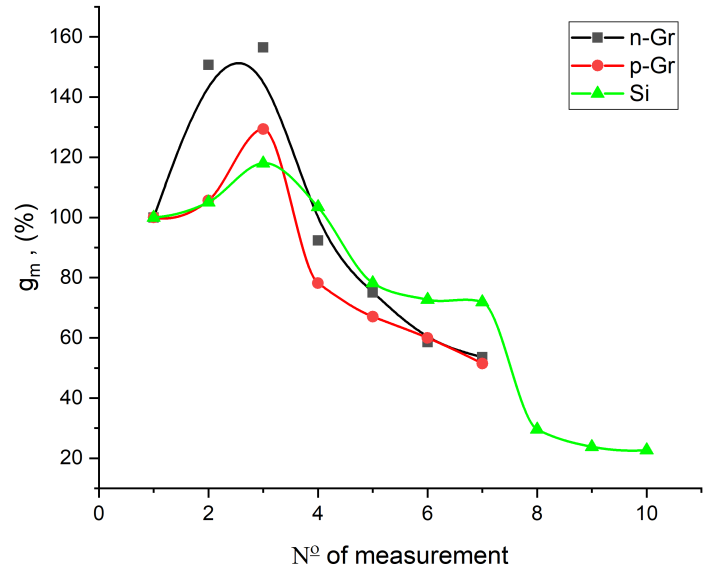

***Supplementary Figure S8.*** *Transconductance degradation versus the number of measurements. Here silicon-like transconductance is color-coded in green, while graphene-like electron and hole transconductance were color-coded in black and red respectively. Results are displayed as a percentage ratio of the transconductance achieved in the first sweep.*

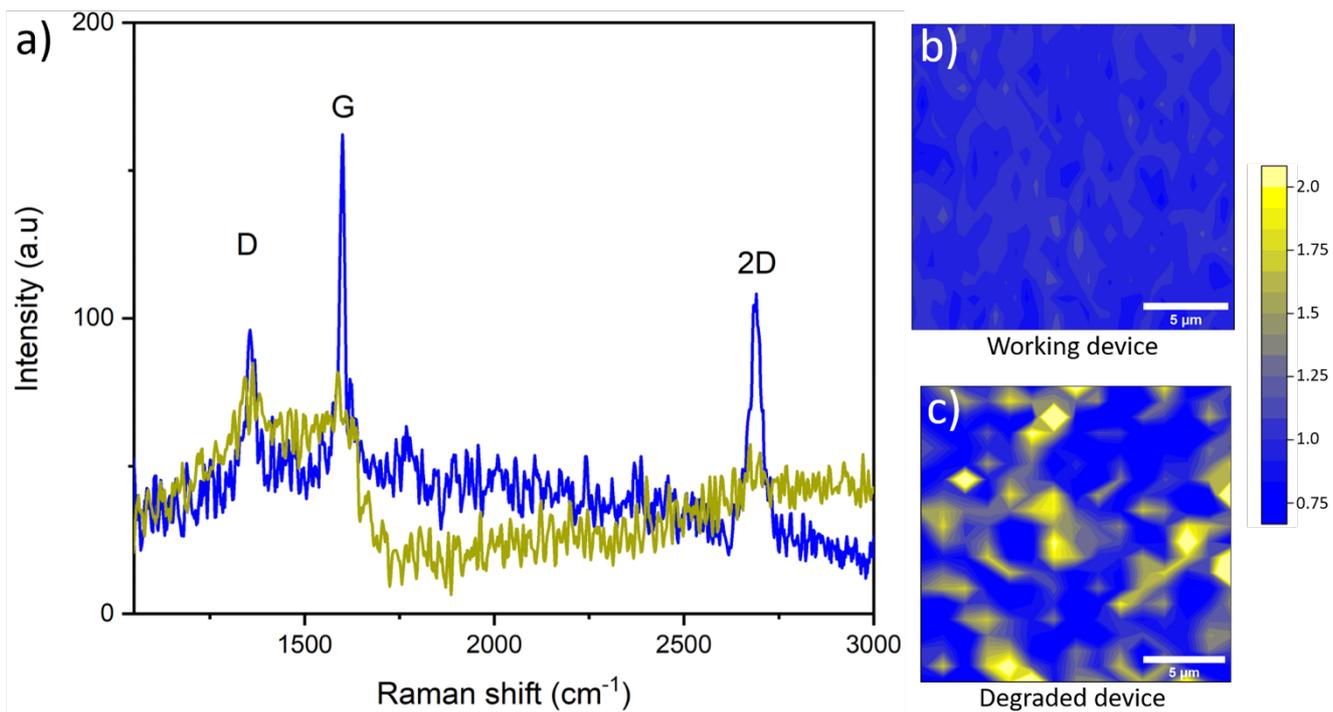

***Supplementary Figure S9.*** Raman spectra of graphene on GoSFETs (a) before (blue curve), and after performance degradation (yellow curve) and corresponding graphene channel footprints (b, c) of the spectra weighting factors. Graphene area is 20μm x20μm with 5μm scale bars on the bottom right corner.

## Supplementary Note S1. Electrostatics of electrolyte-gap-graphene-dipole layer-silicon heterostructure

The electrostatics of the electrolyte-hydrophobic gap-graphene-dipole layer-silicon heterostructure shown in *Fig.3b* is addressed in *Eq.(1)*. To solve it, the following explicit equations are used for the calculation of the channel charge density and capacitance:

**Graphene** [1]–[3]

Charge density:
$$Q_{net}[V_c] = \frac{Kk_1}{2}\left(V_c\sqrt{1+\left(\frac{V_c}{k_1}\right)^2} + k_1 \sinh^{-1}\left[\frac{V_c}{k_1}\right]\right) \quad (1)$$

$$K = \frac{2q^3}{\pi(\hbar v_F)^2}$$

$$k_1 = k_B T \ln[4]$$

where $q$ is the elementary charge; $v_F \approx 10^6$ m/s is the Fermi velocity; $\hbar$ is the reduced Planck constant; $T$ is the temperature and $k_B$ is the Boltzmann constant.

Quantum capacitance:
$$C_q[V_c] = \frac{\partial Q_{net}}{\partial V_c} = \frac{2q^2 k_B T \ln[4]}{\pi(\hbar v_F)^2}\sqrt{1+\left(\frac{qV_c}{k_B T \ln[4]}\right)^2} \quad (2)$$

**Silicon** [4], [5]

Charge density: $Q_{si}[\psi_s,\phi_n] = -\varepsilon_{si} E[\psi_s,\phi_n]$

**n$^+$–p–n$^+$ GoSFET**
$$E[\psi_s,\phi_n] = \pm\sqrt{\frac{2qk_B T N_A}{\varepsilon_{si}}\left(\left(e^{-\frac{q\psi_s}{k_B T}}+\frac{q\psi_s}{k_B T}-1\right)+\frac{n_i^2}{N_A^2}\left(e^{-\frac{q\phi_n}{k_B T}}\left(e^{\frac{q\psi_s}{k_B T}}-1\right)-\frac{q\psi_s}{k_B T}\right)\right)}$$

**p$^+$–p–p$^+$ GoSFET**
$$E[\psi_s,\phi_p] = \pm\sqrt{\frac{2qk_B T N_A}{\varepsilon_{si}}\left(e^{\frac{q\phi_p}{k_B T}}\left(e^{-\frac{q\psi_s}{k_B T}}+\frac{q\psi_s}{k_B T}-1\right)+\frac{n_i^2}{N_A^2}\left(e^{\frac{q\psi_s}{k_B T}}-\frac{q\psi_s}{k_B T}-1\right)\right)} \quad (3)$$

where the positive (negative) sign applies when $\psi_s > 0$ ($< 0$) for a *p*-type semiconductor. $\varepsilon_{si} = 11.7\varepsilon_0$ is the silicon dielectric permittivity. $q\phi_n$ ($q\phi_p$) stands for the quasi-Fermi level in the silicon channel in the inversion (accumulation) GoSFET.

Silicon capacitance: $C_{si}[\psi_s,\phi] = -\frac{\partial Q_{si}}{\partial \psi_s}$

**n$^+$–p–n$^+$ GoSFET**
$$C_{si}[\psi_s,\phi_n] = \pm\sqrt{\frac{2q\varepsilon_{si}N_A}{k_B T}}\frac{\left(-e^{-\frac{q\psi_s}{k_B T}}+1\right)+\frac{n_i^2}{N_A^2}\left(e^{\frac{q(\psi_s-\phi_n)}{k_B T}}-1\right)}{2\sqrt{\left(e^{-\frac{q\psi_s}{k_B T}}+\frac{q\psi_s}{k_B T}-1\right)+\frac{n_i^2}{N_A^2}\left(e^{-\frac{q\phi_n}{k_B T}}\left(e^{\frac{q\psi_s}{k_B T}}-1\right)-\frac{q\psi_s}{k_B T}\right)}}$$

**p$^+$–p–p$^+$ GoSFET**
$$C_{si}[\psi_s,\phi_p] = \pm\sqrt{\frac{2q\varepsilon_{si}N_A}{k_B T}}\frac{e^{\frac{q\phi_p}{k_B T}}\left(-e^{-\frac{q\psi_s}{k_B T}}+1\right)+\frac{n_i^2}{N_A^2}\left(e^{\frac{q\psi_s}{k_B T}}-1\right)}{2\sqrt{e^{\frac{q\phi_p}{k_B T}}\left(e^{-\frac{q\psi_s}{k_B T}}+\frac{q\psi_s}{k_B T}-1\right)+\frac{n_i^2}{N_A^2}\left(e^{\frac{q\psi_s}{k_B T}}-\frac{q\psi_s}{k_B T}-1\right)}} \quad (4)$$

## Supplementary Note S2. Drift-diffusion transport in GoSFETs

The drift-diffusion theory has been adopted to describe the carrier transport through the graphene and silicon channels. The current is calculated as $I_{DS} = WQ_t(x)\mu(x)dV(x)/dx$ from the mobile charge density and closed-form expressions are collected as follows:

**Graphene** [1]–[3]

The ambipolar mobile charge density in graphene can be expressed as a quadratic polynomial given that symmetrical electron and hole mobilities are considered [3], [6]:

<u>Mobile charge density</u>:
$$Q_{tot}[V_c] = \frac{K}{2}(V_c^2 + \alpha) \tag{5}$$
$$\alpha = \frac{(\pi k_B T)^2}{3q^2} + \frac{2}{K}\sigma_{pud}$$
$$\sigma_{pud} = \frac{q\Delta^2}{\pi(\hbar v_F)^2}$$

where a residual charge density due to electron-hole puddles [7], $\sigma_{pud}$, has been considered, with $\Delta$ being the inhomogeneity of the electrostatic potential.

<u>Drain current</u>:

$$I_{DS}[V_{cs}, V_{cd}] = \mu_g \frac{W_g}{L_{eff}[V_{cs}, V_{cd}]} \left( \frac{K}{2} \left( \frac{V_c^3}{3} + k_1 V_c + \frac{\sqrt{1 + \left(\frac{V_c}{k_1}\right)^2}(k_1^2 + 2V_c^2 + 4\alpha) - Kk_1^2(k_1^2 - 4\alpha)\sinh^{-1}\left[\frac{V_c}{k_1}\right]}{8C_z} \right) \right) \Bigg|_{V_{cs}}^{V_{cd}}$$

$$L_{eff}[V_{cs}, V_{cd}] = L + \frac{\pi \mu_g}{2C_z v_F}(V_{cd} - V_{cs} + Q_{net}[V_{cd}] - Q_{net}[V_{cs}]) \tag{6}$$
$$C_z = C_{dip} + C_{int}$$
$$C_{int} = (C_{DL}^{-1} + C_{Stern}^{-1} + C_{gap}^{-1})^{-1}$$

where $V_{cs}$ ($V_{cd}$) is the graphene potential at the source (drain) edge and it is calculated from *Eq.(1)* when $V(x=0)=V_S$ ($V(x=L)=V_D$).

**Silicon** [4], [5]

The unipolar mobile charge density in silicon is the inversion charge density, $Q_i$, in the n$^+$–p–n$^+$ GoSFET, while the depletion charge density, $Q_d$, is considered in the case of the accumulation p$^+$–p–p$^+$ counterpart, with $Q_{si} = Q_i + Q_d$.

**n$^+$–p–n$^+$ GoSFET** (charge-sheet model approximation [5])

<u>Inversion charge density</u>:
$$Q_i[\psi_s, \phi_n] = Q_{si}[\psi_s, \phi_n] - Q_d[\psi_s] \tag{7}$$
$$Q_d[\psi_s] = -\sqrt{2\varepsilon_{si}qN_A\psi_s}$$

Drain current:
$$I_{DS}[\psi_{s,s}, \psi_{s,d}] = \mu_{e,Si}\frac{W_s}{L}\int_{\psi_{s,s}}^{\psi_{s,d}}(-Q_i[\psi_s, \phi_n])\frac{d\phi_n}{d\psi_s}d\psi_s$$

$$\frac{d\phi_n}{d\psi_s} = 1 + \frac{k_BT}{q}\frac{C_{dip}Q_{si}[\psi_s,\phi_n] + \varepsilon_{si}qN_A}{Q_{si}[\psi_s,\phi_n]^2 - 2\varepsilon_{si}qN_A\psi_s} \tag{8}$$

$$I_{DS}\begin{bmatrix}\psi_{s,s},\psi_{s,d}\\V_s,V_d\end{bmatrix} = \mu_{e,Si}\frac{W_s}{L}\left(\psi_s\left((-Q_{si}[\psi_s,\phi_n]) + C_{dip}(\psi_s + k_BT)\right) - \frac{C_{dip}\psi_s^2}{2} - \frac{2\sqrt{2\varepsilon_{si}qN_A}\psi_s^{\frac{3}{2}}}{3} + k_BT\sqrt{2\varepsilon_{si}qN_A\psi_s}\right)\Bigg|_{\substack{\psi_{s,s}\\V_s}}^{\substack{\psi_{s,d}\\V_d}} \tag{9}$$

where $\psi_{s,s}$ ($\psi_{s,d}$) is the surface potential at the source (drain) edge and it is calculated from *Eq.(1)* when $V(x=0)=V_S$ ($V(x=L)=V_D$). $\phi_n$ is the quasi-Fermi potential in the silicon channel and satisfies the boundary condition $\phi_n(x=0)=V_S$ ($\phi_n(x=L)=V_D$). In order to obtain *Eqs. (8)-(10), Eq. (3)* has been reduced by considering the terms that are significant in depletion and inversion where $q\psi_s/(k_BT) \gg 1$ [5].

**p$^+$–p–p$^+$ GoSFET**

Depletion charge density: $\quad Q_d[\psi_s, \phi_p] \approx Q_{si}[\psi_s, \phi_p]$ \hfill (10)

Drain current: $\quad I_{DS}[\psi_{s,s}, \psi_{s,d}] = \mu_{h,Si}\frac{W_s}{L}\int_{\psi_{s,s}}^{\psi_{s,d}}Q_d[\psi_s, \phi_p]\frac{d\phi_p}{d\psi_s}d\psi_s$

$$\frac{d\phi_p}{d\psi_s} = \frac{2k_BTC_{dip}}{qQ_{si}[\psi_s,\phi_p]} + \frac{1}{1-e^{q\psi_s/k_BT}} \tag{11}$$

$$I_{DS}[\psi_{s,s},\psi_{s,d}] = \mu_{h,Si}\frac{W_s}{L}\left[\left(\frac{2k_BTC_{dip}}{q}\psi_s\right)\Bigg|_{\psi_{s,s}}^{\psi_{s,d}} + \int_{\psi_{s,s}}^{\psi_{s,d}}\frac{Q_{si}[\psi_s,\phi_p]}{1-e^{q\psi_s/k_BT}}d\psi_s\right] \tag{12}$$

where $\psi_{s,s}$ ($\psi_{s,d}$) is the surface potential at the source (drain) edge and it is calculated *Eq.(1)* when $V(x=0)=V_S$ ($V(x=L)=V_D$). $\phi_p$ is the quasi-Fermi potential in the silicon channel and satisfies the boundary condition $\phi_p(x=0)=V_S$ ($\phi_p(x=L)=V_D$). In order to get *Eqs. (11)-(12), Eq. (3)* has been reduced by considering the terms that are significant in depletion and inversion where $q\psi_s/(k_BT) \gg 1$. The integral in *Eq. (12)* is numerically solved.

# Supplementary Note S3. Study of the charge and capacitance in the accumulation GoSFET

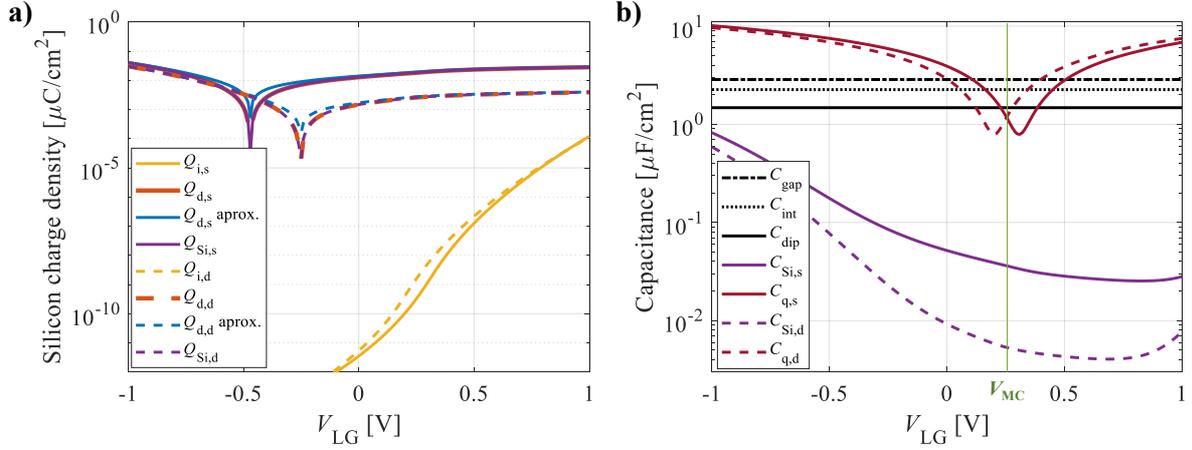

***Supplementary Figure S10 a)** Inversion, depletion and total charge density in the silicon channel; and **b)** relevant capacitances of the accumulation GoSFET (interfacial, gap, dipole, silicon and graphene quantum capacitances) at the drain and source edges as a function of the gate voltage at a drain bias $V_{DS} = -0.1V$.*

The charge density in the silicon channel is depicted in *Fig. S10a* and it is split into the depletion and inversion charge densities that have been numerically calculated as follows [4]:

$$Q_n[\psi_s, \phi_p] = q \frac{n_i^2}{N_A} \int_0^{\psi_s} \frac{e^{q\psi/k_BT} - 1}{E[\psi, \phi_p]} d\psi$$
$$Q_d[\psi_s, \phi_p] = -qN_A \int_0^{\psi_s} \frac{1 - e^{q(\phi_p - \psi)/k_BT}}{E[\psi, \phi_p]} d\psi \quad (13)$$
$$Q_{Si}[\psi_s, \phi_p] = Q_n[\psi_s, \phi_p] + Q_d[\psi_s, \phi_p] = -\varepsilon_{si} E[\psi_s, \phi_n]$$

where $E[\psi, \phi_p]$ is written in *Eq. (3)*. As can be observed in *Fig. S10a*, within the gate voltage window under test, the depletion charge density can be approximated by the total charge density, since the inversion charge density is negligible, $Q_d[\psi_s, \phi_p] \approx Q_{si}[\psi_s, \phi_p]$. In addition, for the drain current equation, the quantity $\frac{d\phi_p}{d\psi_s}$ needs to be calculated from the electrostatics shown in *Eq.(1)*. For such a purpose, *Eq. (11)* and ultimately *Eq. (12)* are obtained by assuming:

$$Q_d[\psi_s, \phi_p] \approx Q_{Si}[\psi_s, \phi_p] = -\varepsilon_{si} E[\psi_s, \phi_n] \approx \sqrt{2qk_BT\varepsilon_{si}N_A \left( e^{\frac{q\phi_p}{k_BT}} \left( e^{-\frac{q\psi_s}{k_BT}} - 1 \right) \right)} \quad (14)$$

where only the terms that are significant in depletion and inversion (where $q|\psi_s|/(k_BT) \gg 1$) have been considered. *Fig. S10a*, shows a comparison of $Q_d[\psi_s, \phi_p]$ calculated either by *Eq. (13)* or *Eq. (14)*, the latter labelled as "aprox.".

A discrepancy between both calculations is produced for $|\psi_s|$ values around $k_B T$, according to *Fig.4d*. In such a situation, the transport charge density is overestimated and therefore so will be the drain current calculation.

On the other hand, *Fig. S10b* shows the relevant capacitances of the hybrid device under study. It can be checked that the equivalent interfacial capacitance is 2.1 by the gap capacitance, $C_{int} = (C_{DL}^{-1} + C_{Stern}^{-1} + C_{gap}^{-1})^{-1} \approx C_{gap}$, given that the $C_{DL}$ and $C_{Stern}$ values are around one order of magnitude higher (not shown in *Fig. S10b*). Regarding the silicon capacitance, we observe an increment for negative values of the gate voltage because of the accumulation of holes at the silicon surface. As the inversion charge density is still low for the gate voltage window considered (*see Fig. S10a*), the increment in the silicon capacitance for positive gate bias values is hardly noticeable. The quantum capacitance of graphene, $C_q$, shows the typical V-shape with the gate voltage [8], [9]. The vertex of the V-shaped curve is produced when the graphene chemical potential is $V_c = 0$, which means that the local Fermi level is at the Dirac energy. In addition, the minimum conductivity voltage $V_{MC}$ has been highlighted where the condition $C_{q,s} = C_{q,d}$ occurs and the graphene channel is half filled with electrons and half filled with holes.

## Supplementary Note S4. Assessment of the n$^+$-p-n$^+$ GoSFET: theory vs. measurements

*Supplementary Table II* collects the electrical and physical parameters used to model the inversion n$^+$-p-n$^+$ GoSFET. It must be highlighted that the parameters used to predict the n$^+$-p-n$^+$ GoSFET are the same that the ones used for the p$^+$-p-p$^+$ GoSFET, collected in Table I, except the values of the mobilities ($\mu_g$ = 40 cm$^2$/Vs; $\mu_{e,Si}$ = 225 cm$^2$/Vs), the residual charge density due to electron-hole puddles ($\sigma_{pud}/q$ = 5.4·10$^{12}$ cm$^{-2}$) and the offset bias ($V_{fb2}$-$V_D$ = -0.41V). The graphene quality of the GoSFET sample under test is better given that the residual charge density is lower and the mobility modestly higher [10].

*Fig. S11a* shows the excellent agreement between the modeling results of the transfer characteristics (TCs) of the inversion GoSFET described in Supplementary Table II and the experimental measurements (forward and backward sweeps). Similarly to its accumulation counterpart, a point of minimum conductivity is observed. We also show in *Fig. S11b* the theoretical current of the n$^+$-p-n$^+$ GoSFET for $V_{DS}$ = -0.2V split into the graphene and silicon current contributions. The current at the graphene channel ($I_{DS,G}$) is modulated with the gate voltage, showing the typical V-shape. The minimum conductivity of the inversion GoSFET occurs at the gate bias labelled as $V_{MC}$ = 0.19V, and it happens for both graphene and silicon currents at the same bias evidencing the large electric coupling between both channels. In *Fig. S11c* the modulation of the graphene electrostatic potential at the drain ($V_D$-$V_{cd}$) and source ($V_S$-$V_{cs}$) edges is shown with a similar behavior to the one commented in the main text.

In the silicon channel, the current ($I_{DS,Si}$) can be split into three regions according to the different operation regions (*Fig. S11b*). To this purpose, in *Fig. S11d*, we show the surface potential at the drain ($\psi_{s,d}$) and source ($\psi_{s,s}$) edges for the inversion GoSFET (solid lines) together with vertical gold dashed lines that delimit the operation regions: deep subthreshold, weak inversion and strond inversion, depending on the conditions $\psi_s < 2\phi_B$, $2\phi_B < \psi_s < \phi_B$, $\psi_s > \phi_B$, respectively. In the subthreshold region, the current is non-negligible according to *Fig. S11b*, contrary to a conventional SiFET. This is caused by the graphene gating effect. The electrostatic graphene potential ($V$-$V_c$) that controls the silicon channel is not uniform along the channel and produces a splitting of the surface potentials at the source and drain edges generating a non-negligible diffusion current. We compare this behavior with that of a typical SiFET, showing the surface potentials of a SiFET (dashed lines) in *Fig. S11d*. In the subthreshold region, the surface potentials of a SiFET are equal, so the charge density is homogeneous along the channel and therefore the diffusion current is negligible. In a GoSFET, the decreasing subthreshold current is compensated with a positive increment of the graphene current as shown in *Fig. S11b*. On the other hand, we observe that the vertex of the V-shaped transfer curve of the GoSFET (purple line) shown in *Fig. S11b* happens at the weak inversion region in the silicon channel. The minimum conductivity is achieved at $V_{MC}$, coinciding with the Dirac voltage and, in addition, the surface potentials in silicon at the drain and source sides get close at this bias, according to *Fig. S11d*, therefore, minimizing the diffusion current. For $V_{LG}$ higher and lower than $V_{MC}$, the surface potentials start to split, so we observe an increment in the current. In particular, for $V_{LG} > V_{MC}$, more electrons are attracted to the surface, and the drift component of the current is expected to increment. For $V_{LG} > 0.41V$, the condition $\psi_s > 2\phi_B$ is accomplished, and the inversion channel is formed with the conduction mechanism similar to a conventional SiFET. In conclusion, contrary to what happens in a conventional SiFET that is controlled by a constant gate potential along the channel, the non-uniform graphene potential ($V-V_c$), that controls the silicon conductivity in a GoSFET, originates a departure of the drain and source surface potentials that produces a relevant diffusion current in the deep subthreshold and weak inversion regions.

SUPPLEMENTARY TABLE II.- N$^+$-P-N$^+$ GOSFET MODEL PARAMETERS

| Graphene | | Silicon | | | |
|---|---|---|---|---|---|
| $L$ (μm) | 20 | $L$ (μm) | 20 | $C_{DL}$ (μF/cm$^2$) | 23 |
| $W_g$ (μm) | 2 | $W_s$ (μm) | 2 | $C_{Stern}$ (μF/cm$^2$) | 20 |
| $V_{fb1}$ (V) | 0.3 | $V_{fb2}$-$V_D$ (V) | -0.41 | $C_{gap}$ (μF/cm$^2$) | 2.86 |
| $\mu_g$ (cm$^2$/Vs) | 40 | $\mu_{e,Si}$ (cm$^2$/Vs) | 225 | $C_{dip}$ (μF/cm$^2$) | 1.48 |
| $\sigma_{pud}/q$ (cm$^{-2}$)* | 5.4×10$^{12}$ | $N_A$ (cm$^{-3}$) | 4×10$^{15}$ | $T$ (K) | 300 |

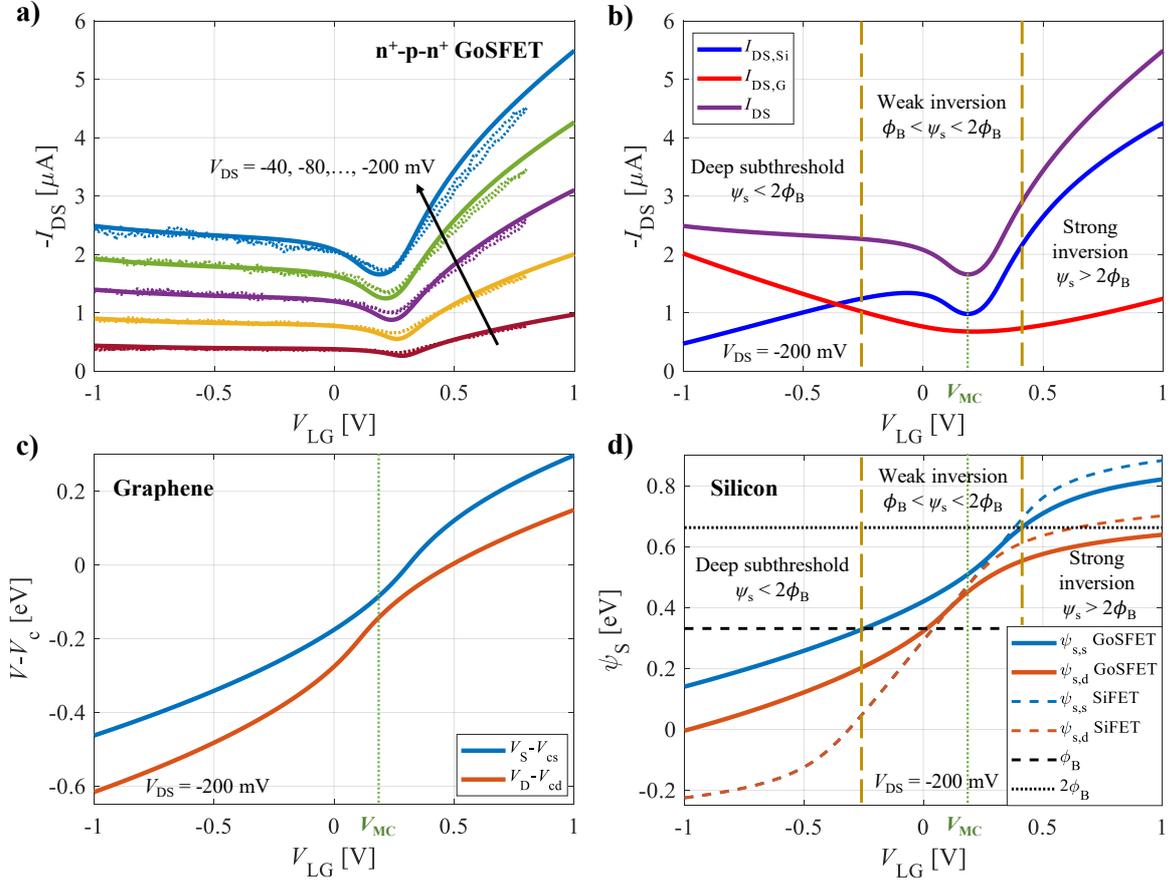

***Supplementary Figure S11. a)*** *Transfer characteristics of the inversion $n^+$-p-$n^+$ GoSFET described in Table II for five different drain biases. Simulations are plotted with solid lines and experimental measured data (forward and backward sweeps) with dots.* ***b)*** *Theoretical drain current of the inversion GoSFET for $V_{DS} = -0.2V$ (purple) split into the graphene (red) and silicon (blue) current contributions. A vertical green dotted line marks the gate bias that achieves the minimum conductivity, labelled as $V_{MC}$.* ***c)*** *Modulation of the graphene electrostatic potential $V-V_c = E_D/q$ at the source (blue) and drain (orange) ends with the gate voltage at $V_{DS} = -0.2V$.* ***d)*** *Modulation of the silicon surface potential at the source (blue) and drain (orange) edges in the inversion GoSFET (solid lines) and a SiFET (dashed lines) with the gate voltage at $V_{DS} = -0.2V$. Vertical gold dashed lines separate the different transistor operation regions.*

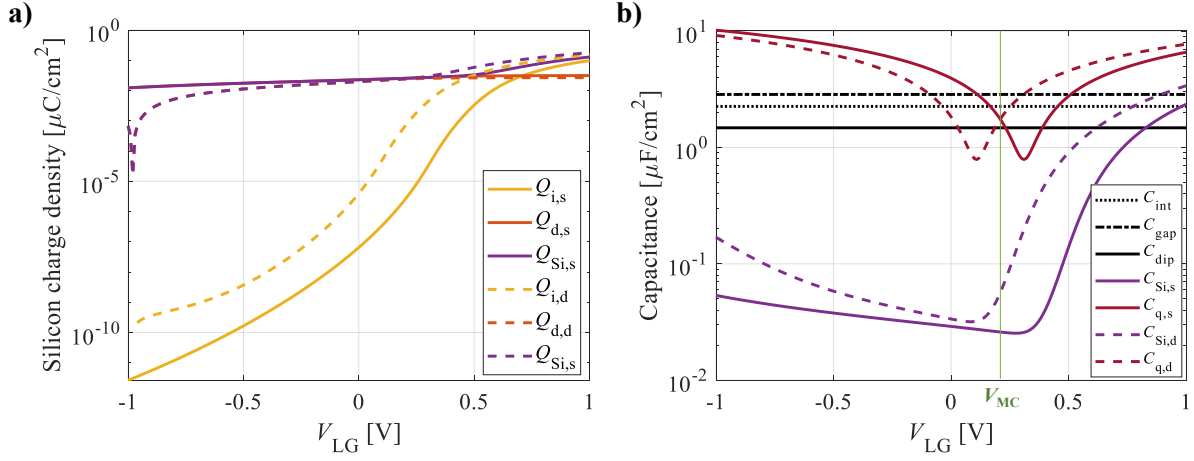

***Supplementary Figure S12. a)*** *Inversion, depletion and total charge density in the silicon channel; and **b)** relevant capacitances of the $n^+$-p-$n^+$ GoSFET (interfacial, gap, dipole, silicon and graphene quantum capacitances) at the drain and source edges as a function of the gate voltage at a drain bias $V_{DS} = -0.2V$.*

The charge density in the silicon channel for the inversion GoSFET is depicted in *Fig. S12a* and it is broken into the depletion and inversion charge densities that have been numerically calculated as follows [4]:

$$Q_n[\psi_s, \phi_n] = q \frac{n_i^2}{N_A} \int_0^{\psi_s} \frac{e^{q(\psi-\phi_n)/k_BT} - 1}{E[\psi, \phi_n]} d\psi$$

$$Q_d[\psi_s, \phi_n] = -qN_A \int_0^{\psi_s} \frac{1 - e^{-q\psi/k_BT}}{E[\psi, \phi_n]} d\psi \qquad (15)$$

$$Q_{Si}[\psi_s, \phi_p] = Q_n[\psi_s, \phi_p] + Q_d[\psi_s, \phi_p] = -\varepsilon_{si}E[\psi_s, \phi_n]$$

where $E[\psi, \phi_n]$ is written in *Eq. (3)*. As can be observed in *Fig. S12a*, the silicon charge density is dominated by the inversion charge in the strong inversion region, when $\psi_s > 2\phi_B$ (cf. *Fig. S11d*). The inversion charge is considered to be the mobile charge density involved in the drain current of the inversion GoSFET. To get an analytical expression for the drain current calculation, again, an approximation of *Eq. (3)* has been assumed by considering only the terms that are significant in depletion and inversion where $q|\psi_s|/(k_BT) \gg 1$, as it is done in [5]. According to *Fig. S11d*, such approximation is accurate within the gate voltage window considered.

On the other hand, *Fig. S12b* shows the relevant capacitances of the inversion GoSFET. The gap, dipole, interfacial and graphene quantum capacitances behave similarly to the ones in the accumulation counterpart shown in *Fig. S10b*. However, the silicon capacitance shows an increment for $V_{LG} > V_{MC}$ due to the formation of the inversion channel at the channel surface. For $V_{LG} < V_{MC}$ the device is in the weak inversion and deep subthreshold region where holes are marginally accumulated at the silicon surface, so we observe an increment in the silicon capacitance.